\documentclass[12pt]{article}

\usepackage{amsmath}
\usepackage{amssymb}
\usepackage{eufrak}

\newcommand{\eps}{\varepsilon}

\begin{document}

\title{Structure of residual interaction in spherical nuclei
\footnote{Zh. Eksp. Teor. Fiz. {\bf 63}, 1961--1977 (1972) [Sov. Pys. JETP
{\bf 36,} No.~6, 1036-1044 (1973)]}}

\author{A.M. Kamchatnov$^{\dagger}$ and V.G. Nosov$^{\ddagger}$\\
$^{\dagger}${\small\it Institute of Spectroscopy, Russian Academy of Sciences, Troitsk, Moscow Region,
142192 Russia}\\
$^{\ddagger}${\small\it Russian Research Center Kurchatov Institute, pl. Kurchatova 1, Moscow,
123182 Russia}
}

\maketitle

\begin{abstract}
The effect of residual interaction between nucleons (quasiparticles) on shell oscillations
of the masses of spherical nuclei is considered. The singularity of the ground state energy
of the system in the vicinity of nucleon magic numbers is analyzed for various types of
the dependence of residual interaction on orbital momentum of the quasiparticle. It is
shown that only the perturbation band width of the Fermi distribution due to residual
interaction which is proportional to the square of the angular momentum vector is consistent
with the character of the magic cusps. The coupling constants between the quasiparticles
are determined on the basis of the available data. The constant decreases rapidly with
increase of nuclear radius. Possible consequences pertaining to the energy spectrum of
infinite nuclear matter are discussed.
\end{abstract}

\section{Introduction}

In a preceding paper [1] we considered the influence of the sharpness of the boundary of
quasiparticle Fermi distribution on the characteristics of spherical nuclei. It turned
out that the number $\widetilde{N}(\rho_f)$ of single-quasiparticle states located below
the boundary $\rho_f=k_fR\gg1$ ($k_f$ is the corresponding value of the wave number and
$R$ is the radius of the nucleus) is strictly speaking not an analytic function of its
argument. In the $\rho$ scale, the singularities of the function are equidistant, with
an interval $\pi/2$; a qualitative idea of their character is gained from Fig.~la. These
are the magic cusp points (jumps of the first derivative); a consistent analysis shows
that singularities of this type are possessed also by the energy of the ground state of
the body, i.e., the mass of the nucleus. Figure 2 shows schematically a typical experimental
plot of the mass in the vicinity of a magic nucleus
\footnote{We emphasize the physically unrealistic character of the "sub-magic" nucleon numbers,
which can be formally set in correspondence with the filling of each $j$-level in a certain
spherically-symmetrical potential well. No phenomenon pertaining to the nucleus as a
whole is observed experimentally in this case; nor does a consistent theoretical analysis
seem to predict any phenomenon. Moreover, not all the characteristics of the referred-to
``well" admit of a clear-cut physical definition. The question is: what takes place at
the point where the magic nucleus is located? Does the Fermi boundary rise because the
occupation of the last $j$-level becomes filled (the most frequently advanced point of
view) or, to the contrary, does the edge of the ``well" drop by a jump in a direction
opposite to the Fermi boundary? This question has apparently no sufficiently distinct
physical meaning, all the more since the position of the bottom of the ``well" is not
a rigorous quantitative concept, owing to the strong damping of the deep quasiparticles.
These jumps of the chemical potential occur by far not after each filling of the next
$j$-level. There are no sub-magic phenomena in spherical nuclei at all, there are only
true magic nuclei. Their position is given theoretically by formulas (19) and (3) of
the preceding paper [1].}.
We have in mind relatively ``weak"
singularities; they appear in the higher-order terms of the expansion in the reciprocal
powers of $\rho_f$. We note that in this approximation, which is of interest in nuclear
physics, there are no grounds whatever for identifying the function $\widetilde{N}(\rho_f)$
with the total number of true particles $N(\rho_f)$. Furthermore, we can advance the
following considerations: The roughly intuitive concept of the quasiparticle as moving
in the average field produced by practically all the particles of the body accounts
nevertheless quite well for the main gist of the phenomenon. Since the dimensionless
parameter $\rho_f = k_fR$ is determined, roughly speaking, from the characteristics of
this average field, it is natural to assume this field to have a smooth dependence on
the total number of particles $N$, as is indeed assumed (see Fig.~1b). It is easy in
practice to make the transition to the physically most significant $N$ scale, for example,
by expanding in reciprocal powers of $\rho_f$ (see [1], formulas (3), (26), and (27) and
the explanations pertaining to them).

We shall now comment on the nature of the shell and magic oscillations in spherical nuclei
from a point of view that reveals clearly the role of the quantum numbers of the individual
quasiparticles. Of particular importance is the conservation of the orbital angular momentum
$l$; the eigenvalues $\rho = kR$ can then be graphically represented by points on the
$(l, \rho)$ plane. It is easily seen from Fig.~3 how they are grouped in the region of
relatively small orbital angular momenta. The ``Regge trajectories" drawn in accordance
with the rule $2n + l = p$ ($n$ is the principal quantum number and $p$ is the number of
the trajectory) lie near their maximum. More concretely, their form is given by the equation
\begin{equation}\label{1}
    \Delta\rho=-(l+1/2)^2/2\rho
\end{equation}
(see also Appendix 1). The vertical distance between the curves is equal to $\pi/2$,
corresponding precisely to the aforementioned interval between the neighboring shells in
the $p$ scale \footnote{ln connection with Fig.~3, there is a curious, simple, and
experimentally confirmed
consequence: each nuclear shell contains either one $s$ or one $p$ state, and the
corresponding single-particle levels lie close in energy to the end of the filling of
this shell. For the atom, there is no such theorem. The absence of any far-reaching
analogy with the shell structure of the atom was already indicated in the preceding
paper (see footnote 2 of [1]). Certain features of the nuclear structure, which are
weakly pronounced in the case of the atom, were mentioned also in footnote 1 above.}.
We can now easily visualize the situation that arises if the Fermi boundary
\begin{equation}\label{2}
    \rho=\rho_f
\end{equation}
is assumed to move, say, upward. After the levels of the last Regge trajectory are exhausted at
the tangency point (see Fig.~3), the density of states $d\widetilde{N}/d\rho$ decreases by jump.
A more formal natural interpretation is that the function $\widetilde{N}(\rho)$ has an
oscillating component $\widetilde{N}_1(\rho)$, the period of which is determined by the
distance between the trajectories in Fig.~3.

This specific grouping of the single-quasiparticle levels, due to the quantum number $l$, is
reflected also in the energy $E$ of the nucleus. Its oscillating component is equal to
\begin{equation}\label{3}
    E_1(\rho_f)=-\varepsilon\widetilde{N}_1(\rho_f),
\end{equation}
where $\widetilde{N}_1(\rho_f)$ is the oscillating part of the number of the states filled
with fermions (quasiparticles). The
proportionality coefficient $-\varepsilon$ is the first functional derivative with respect to
the variation of the distribution function near the Fermi boundary; in other words,
it is the value assumed here by the energy of one quasiparticle. Since, on the other
hand, $\varepsilon = -dE/dN$ is the binding energy of the nucleon in the nucleus, we are
thus dealing with a quantity that can be directly determined from experiment. For case (2)
of an absolutely sharp Fermi boundary, the form of the function $\widetilde{N}_1(\rho_f)$
was determined in [1]. Of particular interest is the decrease $\Delta\varepsilon$ in the
binding energy of the nucleon which is observed in the vicinity of the magic nucleus.
In the ideal case (2) we have
\begin{equation}\label{4}
    (\Delta\varepsilon)_0=\bar{\varepsilon}\frac{d\rho_f^2}{dN},
\end{equation}
where $\bar{\varepsilon}$ is the arithmetic mean of the values of $\varepsilon$ on both sides
of the magic nucleus.

To attempt to make clearer the correspondence with the level distribution with respect to
the momenta, shown in Fig.~3, let us consider the following example, which incidentally is
somewhat artificial: let the quasiparticles fill only all the vacancies that lie below the
boundary, the equation for which is
\begin{equation}\label{5}
    \rho(l)=\rho_f-g(l+1/2)^2/2\rho_f.
\end{equation}
Then calculations similar to those in [1] yield
\begin{equation}\label{6}
    \Delta\varepsilon=\bar{\varepsilon}\frac{d\rho_f^2}{dN}\frac1{1-g}.
\end{equation}
Thus, as $g\to 1$, when the curvilinear boundary (5) of the statistical distribution
coincides with the successive trajectories in Fig.~3 (see also Eq. (1)), the oscillations
tend to infinity in the considered approximation.

The example (5) illustrates only the large sensitivity of the oscillations to redistributions
of the quasiparticles with respect to the quantum states; it has, of course, little in
common with the real situation in spherical nuclei. Finite nuclear dimensions mean apparently
that it is impossible to have a canonical transformation to quasiparticles with respect to
which the case (2) of an ideally abrupt ``stepwise" Fermi distribution would be realized.
This circumstance is frequently called in nuclear physics the ``residual interaction" between
the nucleons. It is felt intuitively that the ``smearing" of the Fermi boundary, due to this
interaction, suppresses, generally speaking, the oscillations. It is seen already from Fig.~3
that as the ``smeared" Fermi boundary moves forward the Regge trajectories, roughly speaking,
are gradually captured. Indeed, in the experiment the magic jumps $\Delta\varepsilon$ are
always smaller than the ideal value $(\Delta\varepsilon)_0$ calculated from formula (4).
However, by far not any smearing of the Fermi boundary is capable of corresponding, even
qualitatively, to the character of the experimental data. For example, a statistical
distribution of the temperature type
\begin{equation}\label{7}
    w_T(\rho)=\frac1{\exp\{(\rho-\rho_f)/\tau\}+1}
\end{equation}
yields for $\widetilde{N}_1(\rho_f)$ an expression that is everywhere differentiable
any number of times. Accordingly, in this case the function $N_1(\rho_f)$ is analytic
and there are no magic singularities~\footnote{The last statement can be regarded also as a consequence
of a theorem of more general character. It is easy to verify that any continuous distribution
of quasiparticles over the states, which does not depend on $l$, results in no singularities
of the type of a jump in the first derivative. In the next two sections of this paper we obtain
a relation compatible with the experimentally observed picture of the magic phenomena between
the width of the smearing of the Fermi distribution and the quantum number $l$.}.

Let us touch also on the question of the most advantageous way of specifying and investigating
the residual interaction. One can imagine, of course, a situation wherein the residual
interaction between quasiparticles is specified in the form of a corresponding Hamiltonian;
one such model example will be calculated in the next section. It is important, however,
not to lose sight of the following circumstance: although the presence of the interaction
causes, strictly speaking, the energy of the individual quasiparticle to be no longer a
definite quantity, this does not influence the applicability of formula (3) in the approximation
of interest to us. Indeed, let the Fermi-distribution smearing, due to the residual interaction,
have a width $\delta\eps$; this can be naturally interpreted as an uncertainty, of the same
order of magnitude, in the quasiparticle energy. On the other hand, in the $\rho$ scale
(see Fig.~3, and also [1]), the quantity $\delta\rho$ characterizing the oscillations is of
the order of unity. Therefore in the region of importance for the oscillations we have
$\delta\eps\sim(d\eps/d\rho)\delta\rho\sim\eps/\rho_f$. Recognizing that $\rho_f\gg1$,
we have
\begin{equation}\label{8}
    \delta\eps\ll\eps,
\end{equation}
i.e., the first factor in the right-hand side of (3) remains definite with sufficient degree
of accuracy. In other words, the redistribution of the quasiparticles over the states, due
to the residual interaction, should be determined in principle from the condition of the
minimum of the energy of the nucleus as a whole. But then the oscillations can be calculated
without taking into account the additional energy of the interaction between the
quasiparticles. This curious feature of the theory will be illustrated by a concrete
example in the next section.

A ``dynamic," so to speak, treatment (meaning that the interaction Hamiltonian between the
quasiparticles is explicitly specified) would be highly ambiguous. Since, in any case, there
is no canonical transformation that leads to an absolutely abrupt Fermi boundary (2), we
apparently have no sufficiently reasonable criterion for a unique choice of the transformation
to a new quasi-static Hamiltonian. In accordance with (8), the residual interaction can be
more adequately described by directly specifying the quasiparticle distribution function with
respect to the states. By regarding this function $w(\rho,l)$ in a certain sense as a primary
concept, we can apparently hope to use successfully its simple single-parameter approximations.
Such a characteristic feature of the phenomenon as the tendency of the residual interaction
to decrease with increasing dimensions of the nucleus is likewise fairly well accounted for
in this case. We shall return to the pertinent questions in the last two sections.

\section{Simplest model of residual interaction}

When choosing a model example, it is desirable to take into account the experimental fact
that the spin of an even-even nucleus in the ground state is equal to zero, and that for odd
nuclei the spin always has the single-particle value within the framework of the shell model
(see, for example, [2]). This suggests the expression
\begin{equation}\label{9}
    H_{int}^j=-G_j\sum_{m,m'>0}a_{m'}^+a_{-m'}^+a_{-m}a_m
\end{equation}
for the interaction between quasiparticles pertaining to the same $j$-level. Here $a_m^+$ and
$a_m$ are the creation and annihilation operators of a quasiparticle with a $z$-projection of
the angular momentum equal to $m$. The Hamiltonian (9) can be diagonalized exactly (see, for
example, [3]); the eigenvalues are given by the well-known formula of Racah and Mottelson
\begin{equation}\label{10}
    E_{int}=-G_jb_j[\Omega_j-(b_j-1)-s_j],
\end{equation}
where $2\Omega_j=2j+1$ is the total number of vacancies, $b_j$ is the number of interacting
pairs at the $j$-level, and $s_j$ is the number of noninteracting quasiparticles (seniority).

It is clear that the smallest energy of the system corresponds to zero seniority, and
consequently for the nucleus as a whole the situation reduces to a minimization of the sum
\begin{equation}\label{11}
    E=\sum_j[2\eps_jb_j-G_jb_j(\Omega_j-b_j+1)]
\end{equation}
over the $j$-levels ($\eps_j$ is the initial value of the quasiparticle energy). The
redistribution of the quasiparticles should be visualized as occurring at a zero variation
of the quantity
\begin{equation}\label{12}
    \widetilde{N}=\sum_j2b_j.
\end{equation}
This additional condition can be easily taken into account by the method of
Lagrange multiplier. Taking also the Pauli principle into account (i.e., the condition
$0\leq b_j\leq\Omega_j$), we get
\begin{equation}\label{13}
    w_j=\frac{b_j}{\Omega_j}=\left\{
    \begin{array}{cc}
    1, & \eps_j-\eps_f<\Delta_-,\\
    \frac{G_j(\Omega_j+1)-2(\eps_j-\eps_f)}{2G_j\Omega_j},&\quad \Delta_-<\eps_j-\eps_f<\Delta_+,\\
    0,& \eps_j-\eps_f>\Delta_+,
    \end{array}
    \right.
\end{equation}
where $\eps_j$ is the chemical potential and $\Delta_{\mp}=\tfrac12G_j(1\mp\Omega_j)$. We present
also the interaction energy (10) corresponding to the equilibrium distribution:
\begin{equation}\label{14}
    E^j_{int}=\left\{
    \begin{array}{cc}
    -G_j\Omega_j, & \eps_j-\eps_f<\Delta_-,\\
    -\frac{G_j^2(\Omega_j+1)^2-4(\eps_j-\eps_f)^2}{4G_j},&\quad \Delta_-<\eps_j-\eps_f<\Delta_+,\\
    0,& \eps_j-\eps_f>\Delta_+.
    \end{array}
    \right.
\end{equation}
We note that the initial expression (10) was the result of simultaneous diagonalization of the
Hamiltonian (9) and the operator of the number of quasiparticles at the $j$-level; then the
quantum numbers $b_j$ are naturally integers. However, in the subsequent derivation of the
distribution (13) it becomes necessary to differentiate already with respect to variables
$b_j$. They are consequently determined with an error of the order of unity. This accuracy
corresponds precisely to the macroscopic character deduced in [1] for the investigated
phenomena. We ultimately replace $\Omega_j\pm1$ by $\Omega_j=j+\tfrac12$; with the same
accuracy, we can replace the total angular momentum $j$ of the quasiparticle by its orbital
momentum $l$ and neglect the spin-orbit interaction
\footnote{The latter is confirmed by an analysis of the dependence of the spin-orbit interaction on
nuclear dimensions (see, for example, [2]), i.e., in final analysis, on $\rho_f=k_fR$. It must
be stipulated, however, that this pertains for the time being to the calculation of the
effects in the $\rho_f$ scale. Subsequently, the spin-orbit interaction is actually taken
into account in a most substantial manner when the final transition is made to the physical
$N$-scale; see [1].}.

We now change to more convenient variables, using a notation close to that of [1] (see also [4]):
\begin{equation}\label{15}
    \Omega_j\cong\rho\sin\widetilde{\beta}\cong\rho_f\widetilde{\beta},\quad
    \widetilde{\beta}=\arcsin(\tilde{l}/\rho),\quad \tilde{l}=l+\tfrac12,\quad
    G_j=\left.\frac{d\eps}{d\rho}\right|_fg_j.
\end{equation}
We have taken into account here the fact that it is the small $\widetilde{\beta}$ that are
significant for the oscillations. In this limit, we assume a power-law dependence of the
coupling constant on $\widetilde{\beta}$:
\begin{equation}\label{16}
    g_j=g\widetilde{\beta}^{k-1}.
\end{equation}
In terms of the variables $\rho$ and $\widetilde{\beta}$ the distribution (13) takes the form
\begin{equation}\label{17}
    w=\left\{
    \begin{array}{cc}
    1, & \rho-\rho_f<-\tfrac12g\rho_f\widetilde{\beta}^k,\\
    \frac12-\frac{\rho-\rho_f}{g\rho_f\widetilde{\beta}^k},&\quad
    -\tfrac12g\rho_f\widetilde{\beta}^k<\rho-\rho_f<\tfrac12g\rho_f\widetilde{\beta}^k,\\
    0,& \rho-\rho_f>\tfrac12g\rho_f\widetilde{\beta}^k.
    \end{array}
    \right.
\end{equation}

Thus, in terms of the distribution function $w(\rho,\tilde{l})$, the model (9)--(11) turns out
to be the simplest of all the conceivable ones. It represents, in fact, a linear interpolation
between regions where the occupation numbers of the quasiparticles assume the limiting values
zero and unity, see Fig.~4. Instead of analyzing the expression for the second variation,
it is easier to verify the stability of the distribution (13), (17) directly, by effecting
all the conceivable transfers of the pair of interacting quasiparticles between the states,
and calculating the corresponding change of the energy of the nucleus. According to Fig.~4,
four types of quasiparticle transfer are permissible: I--II, I--III, II--II, and II--III.
In all cases we have for the change of the system energy
\begin{equation}\label{18}
    \Delta E\geq G_j^{in}+G_j^{fin},
\end{equation}
i.e., the distribution of the quasiparticles over the states is stable.
To calculate the oscillating part of the nuclear energy, we use the result obtained in [1]:
\begin{equation}\label{19}
    E_1=\sum_{\nu=1}^{\infty}(E^\nu+E^{\nu *}),\quad E^\nu=\iint_0^\infty
    e^{2\pi i\nu(2n+l)}E(n,l)\,dndl,
\end{equation}
where
\begin{equation}\label{20}
    E(n,l)=4\eps_f\tilde{l}w=-4\eps\rho w\sin\widetilde{\beta}
    \cong-4\eps\rho_f\widetilde{\beta}w
\end{equation}
is the energy of the quasiparticles having the corresponding values of the quantum
numbers; the additional doubling is due to the spin of the individual quasiparticles, to which
an energy value $\eps_f=-\eps$ is assigned (we shall discuss the validity of the last
assumption at the end of the section). According to [1] (see also [4])
\begin{equation}\label{21}
    2\pi(2n+l)\cong-4\pi+4\rho+2\rho\widetilde{\beta}^2,\qquad
    dndl=\frac1\pi\rho d\rho\cos^2\widetilde{\beta}d\widetilde{\beta}\cong
    \frac{\rho_f}{\pi}d\rho d\widetilde{\beta}.
\end{equation}
Taking also (17) into account, substituting in (19), and performing the simple integration
with respect to $\rho$, we obtain
\begin{equation}\label{22}
    E^\nu=\frac{\eps\rho_f\exp(4i\nu\rho_f)}{4\pi g\nu^2}\int_0^{\infty}\exp(2i\nu\rho_f
    \widetilde{\beta}^2)\left(\exp(2i\nu g\rho_f\widetilde{\beta}^k)-
    \exp(-2i\nu g\rho_f\widetilde{\beta}^k)\right)\frac{d\widetilde{\beta}}{\widetilde{\beta}^{k-1}}.
\end{equation}

To determine which values of the exponent $k$ are compatible with the character of the
experimental data, it is easiest to turn to the limiting case when the convergence of
the integral (22) is due principally to that term of the argument of the exponential
which is proportional to $g$
\footnote{Strictly speaking, however, the coupling constant can likewise not be exceedingly large,
for then the situation reduces to effects due to individual nucleons at the $s$ or $p$ levels;
see Figs.~3 and 4. This would actually be manifest by a complete disappearance of the magic numbers,
and in any case the macroscopic theory of shell and magic phenomena developed in [1] and in
the present paper would no longer hold. As a result we obtain the requirement $g \ll\rho_f^{k-1}$.}.
We expand $\exp(2i\nu\rho_f\widetilde{\beta}^2)$ in a series, confine ourselves to the
first two terms, and substitute in (19); this yields
\begin{equation}\label{23}
\begin{split}
    E_1\cong-\frac{\eps\rho_f}{\pi kg}\Bigg\{\frac{(2g\rho_f)^{2-4/k}}{g}\int_0^\infty
    \frac{\sin y\,dy}{y^{2-4/k}}\sum_{\nu=1}^\infty\frac{\cos4\nu\rho_f}{\nu^{4/k}}\\+
    (2g\rho_f)^{1-2/k}\int_0^\infty\frac{\cos y\,dy}{y^{2-2/k}}\sum_{\nu=1}^\infty
    \frac{\sin4\nu\rho_f}{\nu^{1+2/k}}\Bigg\},\quad g\gg\rho_f^{k/2-1}.
    \end{split}
\end{equation}

The singularities of (23) are governed by trigonometric series that can be easily
investigated. A singularity of the type of a finite jump in the derivative $dE_1/d\rho_f$
(see Figs.~2 and 1b) can be obtained only from an even series in $t = 4\rho_f - 2\pi p$
with cosines. When $k< 2$ we have $4/k > 2$, and the series of the derivatives diverges
uniformly in accordance with the Weierstrass criterion. Thus, there are no singularities
of this type if $k < 2$.
\footnote{The case $k = 0$ calls for a special analysis. Although the expression under the
summation sign becomes somewhat more complicated in this case, the absence of a jump
in the first derivative of $E$, can be easily proved (see Appendix~2). We note that
the choice $k = 0$ is, in a certain sense, of special interest, since it corresponds
to a Fermi-distribution smearing band whose width is independent of the orbital momentum
(this case was already mentioned in the introduction; see footnote 3). For example, in the
Cooper phenomenon, the angular momentum $l$ of the quasiparticle with respect to the
geometric center would not play any special role, and only the angular momenta of the
components of the Cooper pair with respect to their common center of gravity would be
of importance. We defer additional comments to the last section.}
To the contrary, if $k > 2$ the sum of the series exhibits stronger singularities.
To verify this, we consider the derivative of the series of interest to us; this
derivative is given by
$$
\sum_{\nu=1}^\infty\frac{\sin\nu t}{\nu^\alpha},\qquad \alpha<1.
$$
Near the singular point $t = 0$ we have $\sin\nu t\sim\nu t$ up to a certain limiting
value $\nu=\tilde{\nu}\sim1/|t|$. Replacing the summation by integration, we obtain
the estimate
\begin{equation}\label{24}
    \sum_{\nu=1}^\infty\frac{\sin\nu t}{\nu^\alpha}\sim t\int_0^{\tilde{\nu}}
    \nu^{1-\alpha}d\nu\sim t\tilde{\nu}^{2-\alpha}\sim\frac{t}{|t|^{2-\alpha}}.
\end{equation}
Consequently, at $k > 2$ we have $\alpha = 4/k - 1 < 1$ and the derivative of the first
term of (23) tends to infinity as $t\to 0$ (the result (24) agrees with rigorous mathematical
theorems; see, for example,[5]). Thus, only
\begin{equation}\label{25}
    k=2
\end{equation}
is compatible with the experimental data.

To get rid of the lower bound $g \gg 1$ on the coupling constant, we substitute (25) in (22)
and integrate; as a result we obtain ultimately
\begin{equation}\label{26}
    E_1(N,Z)=-\frac{\varepsilon\rho_f}{2\pi}\left\{f_1(g)\EuFrak{ M}(\rho_f)
    +f_2(g)\EuFrak{ N}(\rho_f)\right\},
\end{equation}
where
\begin{equation}\label{27}
    f_1(g)=\frac1{2g}\ln\left|\frac{1+g}{1-g}\right|,\quad f_2(g)=\frac{\pi}{2g}\theta(g-1),
\end{equation}
$\theta(x) = 1$ at $x > 0$ and $\theta(x) = 0$ at $x < 0$. Plots of the functions $f_1$ and
$f_2$ are given in Fig.~5. The trigonometric series
$$
    \EuFrak{M}(\rho_f)=\sum_{\nu=1}^{\infty}\frac{\cos4\nu\rho_f}{\nu^2}=
    \frac{\pi^2}6-2\pi\left|\rho_f-\frac\pi2p\right|+4\left(\rho_f-\frac\pi2p
    \right)^2,\quad
    \frac\pi2(p-1)<\rho_f<\frac\pi2(p+1)
$$
\begin{equation}\label{28}
    \EuFrak{N}(\rho_f)=\sum_{\nu=1}^{\infty}\frac{\sin4\nu\rho_f}{\nu^2}
\end{equation}
describe shell oscillations in spherical nuclei in the presence of a residual interaction
\footnote{Greatest interest, however, attaches nevertheless to the nearest vicinity of the magic
nucleus, in which the function $\EuFrak{N}$ tends to zero like $t\ln |t|$. At the present
time it is still not quite clear whether allowance for the small term proportional to $\EuFrak{N}$
is an exaggeration of the macroscopic accuracy with which the entire developed concept holds
(see also [1]). This has no effect whatever on the subsequent results pertaining to the magic
jump $\Delta\eps$ of the nucleon binding energy.}.
Differentiating (26), we obtain the discontinuity
\begin{equation}\label{29}
    \Delta\eps=\bar{\eps}\frac{d\rho_f^2}{dN}f_1(g)
\end{equation}
of the nucleon binding energy in the nucleus. According to (27), this discontinuity increases
without limit as $g\to 1$. The physical qualitative picture was already explained in the
introduction and illustrated in Fig.~3. Now, however, unlike in (6), the oscillations tend
to infinity much less rapidly, namely, logarithmically. The weakening of the singularity is
due to the presence of region II, which is not free of fermions (see Fig.~4); on the boundary
between regions I and II, the occupation numbers (17) do not change jumpwise in the considered model.

It is quite doubtful whether such a sharply delineated boundary of the region of intermediate
values of occupation numbers actually exists, and actually the coefficient of the function
$\EuFrak{M}(\rho_f)$ should not become infinite anywhere. We note in this connection the
curious possibility of constructing a simple interpolation formula for the function $f_1(g)$.
Since the transition $g\to 0$ to the case (4) where there is no residual interaction yields
$f_1\to 1$, and since in the asymptotic region $g\gg 1$ we have according to (27)
$f_1\cong 1/g^2$, the following interpolation comes to mind:
\begin{equation}\label{30}
    f_1(g)\approx 1/(1+g^2).
\end{equation}
We then have, in particular,
\begin{equation}\label{31}
    \omega\equiv\frac{(\Delta\eps)_0}{\Delta\eps}-1=\frac1{f_1}-1\approx g^2.
\end{equation}
Thus, the characteristic $\omega$ of the deviation from the ideal case (2) and (4), which
was introduced earlier in [1] from intuitive and empirical considerations, turns out to be
directly expressed in terms of the constant of the residual interaction between the
quasiparticles. Nevertheless, it seems more natural to start from the very beginning from
analytic or nearly-analytic expressions for the distribution function $w(\rho, \widetilde{\beta})$.
One such example will be considered in the next section.

The prospects for choosing the most successful interpolation may become clearer if we turn
finally to the assumption (made above in connection with formula (20)) that each quasiparticle
has a definite energy $-\eps$. There exists also the energy (14) of the interaction between
the quasiparticles. When the spectral component $E_{int}^\nu$ of the energy of this
interaction is calculated by means of formula (19), integration over the region I will
certainly not yield a macroscopic contribution. Indeed (see formula (14) and the text
immediately following it) the only inaccuracy in the determination of $b_j$ corresponds
precisely to the error in the interaction energy $\sim G_j\Omega_j$. We therefore confine
ourselves to integration over the interaction region II, and simple calculations lead to
\begin{equation}\label{32}
    E_{int}^\nu=\left.\frac{d\eps}{d\rho}\right|_f\frac{\rho_f}{16\pi}e^{4i\nu\rho_f}
    \frac{i}{\nu^3}\left\{\frac1{1-g^2}-f_1(g)+if_2(g)\right\}.
\end{equation}
Since $d\eps/d\rho\sim\eps/\rho_f$, the contribution (32) to the energy of the nucleus
is not macroscopic; according to the criteria analyzed in [1], it need not be taken
into account.

This circumstance is apparently not accidental, and is by no way a distinguishing feature
of just this model. From a more general and physically more lucid point of view, this
question was already analyzed in the introduction above.

\section{Case of analytic distribution function}

The fact that the contribution of the interaction energy to the oscillations is of no
importance (at a specified quasiparticle distribution function) enables us to reformulate
the problem in the manner outlined in the introduction, namely, the oscillating part of
the ground-state energy of a spherical nucleus is defined by the function $w$ (which is
the diagonal part of the density matrix relative to the basis of the quasiparticles in
question). The form of the function $w(\rho,\widetilde{\beta})$ is by far not as arbitrary
as it might seem at first glance. First, it must tend rapidly to zero and unity in the
asymptotic regions; it is not very likely that this function is non-monotonic. Further,
the analysis in the preceding section (see in particular (25) and (26)) suggests that only
a quadratic dependence of the width of the Fermi-distribution smearing band on the orbital
angular momentum can be reconciled with the experimentally observed character of the magic
phenomena at more or less arbitrary intensities of the residual interaction. Indeed, it is
easy to show that for all distributions of the type $w((\rho - \rho_f)/\widetilde{\beta}^2)$
the oscillating part of the energy $E_1$ always reduces to a linear combination of the
expression $\EuFrak{M}(\rho_f)$ and $\EuFrak{N}(\rho_f)$. It is therefore natural to attempt
to use a function of the type of the ordinary Fermi distribution (7), but with a modulus
that depends quadratically on the angle $\widetilde{\beta}$:
\begin{equation}\label{33}
    w(\rho,\widetilde{\beta})=\frac1{\exp\left(\frac{\rho-\rho_f}{\tau\widetilde{\beta}^2}\right)
    +1}.
\end{equation}
We calculate the spectral component
\begin{equation}\label{34}
    \widetilde{N}^\nu=\frac4\pi\rho_f^2e^{4i\nu\rho_f}\int_0^\infty d\widetilde{\beta}\widetilde{\beta}
    \exp\{2i\nu\rho_f\widetilde{\beta}^2\}\int w(\rho,\widetilde{\beta})\exp\{4i\nu(\rho-\rho_f)\}d\rho
\end{equation}
of the number of states filled with quasiparticles, and in accord with
\begin{equation}\label{35}
    \widetilde{N}_1=\sum_{\nu=1}^\infty(\widetilde{N}^\nu+\widetilde{N}^{\nu *})
\end{equation}
we use formula (3) (taking into account the conclusions drawn above concerning the role of
the energy of interaction between quasiparticles, this is equivalent to formulas (19)):
\begin{equation}\label{36}
    E_1=-\frac{\eps\rho_f}{4\pi^2g}\sum_{\nu=1}^\infty\left\{\frac{e^{4i\nu\rho_f}}{i\nu^2}
    \zeta\left(2,\frac12-\frac{i}{\pi g}\right)+ \mathrm{c.c.}\right\}.
\end{equation}
We have put here $\tau=(1/4)g\rho_f$ (the convenience of such a change of notation will become clear
later on). Separating in the Riemann $\zeta$ function
\begin{equation}\label{37}
    \zeta(z,q)=\frac1{\Gamma(z)}\int_0^\infty\frac{u^{z-1}e^{-qu}}{1-e^{-u}}du=\sum_{\mu=0}^\infty
    \frac1{(q+\mu)^z}
\end{equation}
the real and imaginary parts, we arrive at the trigonometric series (28):
\begin{equation}\label{38}
    \begin{split}
    E_1(N,Z)&=-\frac{\varepsilon\rho_f}{2\pi}\left\{F_1(g)\EuFrak{ M}(\rho_f)
    +F_2(g)\EuFrak{ N}(\rho_f)\right\},\\
    F_1(g)&=\frac1{2\pi g}\int_0^\infty\frac{\sin(u/\pi g)}{\sinh(u/2)}udu,\\
    F_2(g)&=\frac1{2\pi g}\int_0^\infty\frac{\cos(u/\pi g)}{\sinh(u/2)}udu=\frac{(\pi/2g)}{\cosh^2(1/g)}.
    \end{split}
\end{equation}
Plots of the functions $F_1(g)$ and $F_2(g)$ are shown in Fig.~6.

Just as for the model example (17) and (25) of the preceding section, in the region $g\gg 1$ of
the strong residual interaction there predominates, generally speaking, the function $\EuFrak{M}(\rho_f)$,
which enters here with a coefficient
\begin{equation}\label{39}
    F_2(g)\cong\pi/2g,\qquad g\gg 1.
\end{equation}
With our choice of the normalization of the coupling constant (see above), Eq.~(39) coincides with the
second formula of (27). Continuing this analogy, we note that in the non-analytic model (23), (27)
there was no term with $\EuFrak{N}(\rho_f)$ at all at $g < 1$. In our case, however, this corresponds
to exponential smallness of this term if the coupling between the quasiparticles is weak:
\begin{equation}\label{40}
    F_2(g)\cong\frac{2\pi}ge^{-2/g},\qquad g\ll 1.
\end{equation}
From the more formal point of view, it is curious to note that the limiting behavior of (40) indicates
the presence of an essential singularity at $g = 0$. One cannot exclude the possibility that this
mathematical property of the limiting case $g\to 0$ of an absolutely sharp Fermi boundary is quite
general.

The magic singularities are due only to the function $\EuFrak{M}(\rho_f)$ (see also footnote 7). The
formulas characterizing these singularities are
\begin{equation}\label{41}
    \Delta\eps=\bar{\eps}\frac{d\rho_f^2}{dN}F_1(g),\qquad \omega=\frac{(\Delta\eps)_0}{\Delta\eps}-1=
    \frac1{F_1}-1.
\end{equation}
As seen from Fig. 6, the function $F_1(g)$ has a maximum in the region of intermediate values of
the coupling constant. Thus, the effect of interest to us does not depend monotonically on the intensity
of the residual interaction\footnote{This non-monotonicity is an interesting feature of the phenomenon,
of which we were unaware during the time of writing of the preceding paper [1]; nor is it revealed
by rough interpolations of the type (30). The increase of the oscillations at intermediate values
of $g$ can be qualitatively understood as a sort of resonant amplification due to the similarity
in the form of the lower boundary of the region of intermediate values of occupation numbers and
the corresponding Regge trajectory on Fig.~3. In the non-analytic model examples (6) and (29)
(see 27)) this resonance was extremely sharply pronounced and caused the effect to become infinite
as $g\to 1$. It can be assumed that in the analytic case, too, the presence of a maximum is not
accidental, and that if the Fermi boundary is not absolutely sharp, $g = 0$, the oscillations
actually reach their maximum swing. For the considered distribution (33), this extremum is given
by $g = 0.55$, $F_{1, max}= 1.20$, and $\omega_{min} = -0.17$.}.

\section{Comparison with experiment}

For the 52 magic nuclei considered in the preceding article [1], we obtained the average values
pertaining to seven experimentally accessible cusps (four neutron and three proton) on the mass
curve. The results of such a reduction are given in Table~I. In accordance with the interpolation
(38) and (31), $\sqrt{\omega}$ is the crudest characteristic of the residual interaction. The
reason why $\omega$ (or $\sqrt{\omega}$) may turn out not to be an appropriate characteristic
of the phenomenon of interest to us are given in footnote 8. The coupling constant $g'$ was
calculated from formulas (27) and (29), and the quantity $g$ corresponding to it in the analytic
case (33) was determined from relations (38) and (41).

With respect to the shell oscillations the dependence of the width of the Fermi-distribution
smearing band on $\widetilde{\beta}$ extends only to the angles
\begin{equation}\label{42}
    2\rho_f\widetilde{\beta}^2\sim1,
\end{equation}
after which the integrals describing them (see (22), (34)), in any case, converge rapidly.
According to (33), the corresponding characteristic width in the $\rho$ scale is
\begin{equation}\label{43}
    \overline{\delta\rho}=\tau\widetilde{\beta}^2=\tfrac14g\rho_f\widetilde{\beta}^2=\tfrac18g.
\end{equation}
We change over to the energy scale:
\begin{equation}\label{44}
\overline{\delta\eps}=\left.\frac{d\eps}{d\rho}\right|_f\overline{\delta\rho}=
\frac{\hbar^2}{R^2}\frac{\rho_f}{m^*}\frac{g}8.
\end{equation}
Here $R$ is the radius of the nucleus. We now assume
\begin{equation}\label{45}
    R=1.2\cdot 10^{-13}\,A^{1/3}\,[\mathrm{cm}]
\end{equation}
and, as an estimate, set the effective mass $m^*$ of the quasiparticles equal to the mass of
the free nucleon. The widths $\overline{\delta\eps}$ in MeV are given in Table I.

For the concrete values of the orbital angular momentum, the width $\overline{\delta\eps}$
of the smearing band turns out to be different, since it depends on $\tilde{l} = l + 1/2$
quadratically. Taking the estimate (42) into account, we confine ourselves to not too large
values of $l$, which are characteristic of the considered shells. The results are given in
Table~II.

All the characteristics of the residual attraction
listed in the table, regardless of the degree of their accuracy and of the choice of scale,
indicate consistently that its intensity decreases rapidly with increasing nuclear dimensions
\footnote{The method of coping with the even-odd mass oscillations used in the reduction of the
experimental data was already described earlier (see [1], footnote 12). We note in this
connection that the change in the number of protons by 2 is a reflection of the Coulomb
energy of the nucleus is reflected in the Coulomb energy of the nucleus and in its
derivative with respect to the number of charged particles. Since in fact an energy of
pure electrostatic origin has no singularity whatever, it is advisable to subtract the
``fictitious" effect produced by it from the observed running discontinuities $\Delta\eps$.
In connection with the introduction of such a correction for the Coulomb interaction,
the values of $\omega$ used in the present paper for the proton magic nuclei (see Table~I)
turn out to be somewhat higher than those previously published (see Table~II and Fig.~4 of
[1]). This correction affects little the conclusions concerning the course and magnitude
of the residual interaction of the protons.}.

\begin{center}
\begin{table}
\begin{tabular}{|c|c|c|c|c|c|c|c|c|c|}
\multicolumn{10}{c}{\bf Table I}\\
\hline
$p$ & $N,\,Z$ & \multicolumn{4}{|c|}
{ Neutron magic nuclei} &
\multicolumn{4}{|c|}{ Proton magic nuclei } \\
 & &
$\sqrt{\omega}$ & $g'$ & $g$ & $\overline{\delta\eps}$ (MeV) &
$\sqrt{\omega}$ & $g'$ & $g$ & $\overline{\delta\eps}$ (MeV) \\
\hline
4  & 28 & 2.4 & 2.7 & 3.3 & 5.2 & 1.8 & 2.2 & 2.6 & 3.8 \\
5  & 50 & 1.6 & 2.0 & 2.3 & 3.3 & 1.5 & 1.9 & 2.2 & 2.6 \\
6  & 82 & 1.4 & 1.8 & 2.0 & 2.6 & 1.0 & 1.5 & 1.6 & 1.6 \\
7  & 126 & 1.1 & 1,6 & 1.7 & 1.9 &    &     &     &      \\
\hline
\end{tabular}
\end{table}
\end{center}

\begin{center}
\begin{table}
\begin{tabular}{|c|c|c|c|c|c|c|c|c|}
\multicolumn{9}{c} {\bf Table II
 (The values of $\delta\eps$ are given in MeV.)}\\
\hline
$l$ & \multicolumn{2}{|c|} {$p=4$} & \multicolumn{2}{|c|} {$p=5$} &
\multicolumn{2}{|c|}{ $p=6$} & \multicolumn{2}{|c|} {$p=7$} \\
 &  $\delta\eps_N$ & $\delta\eps_Z$ & $\delta\eps_N$ &
 $\delta\eps_Z$ & $\delta\eps_N$ & $\delta\eps_Z$ & $\delta\eps_N$&
 $\delta\eps_Z$ \\
 \hline
0 &  0.4 & 0.3 &      &     &    0.14 &   0.08 &      &     \\
1 &      &     &  1.9 & 1.5 &         &        & 0.8  &     \\
2 &  10.2& 7.5 &      &     &    3.5  &   2.1  &      &    \\
3 &      &     &  10.2& 8.1 &         &        & 4.3  &    \\
\hline
\end{tabular}
\end{table}
\end{center}

\section{Conclusion and discussion}

The most interesting result, in our opinion, is the dependence of the residual interaction
on the orbital angular momentum of the quasiparticle:
\begin{equation}\label{46}
    \delta\eps\propto \mathbf{l}^2\simeq(l+1/2)^2.
\end{equation}
On the one hand, this result is quite natural. Since we are dealing with a scalar effect,
it should be expressed precisely through the scalar square of the angular-momentum vector.

According to the estimate (42), the angular-momentum values that play an important role
are relatively small. From this point of view, (46) can be regarded as the first term of
a series expansion in powers of the ratio $\tilde{l}/\rho_f$. Why does the expansion have
no zeroth term independent of $l$? Does this mean that in a nucleus of finite dimensions
the residual interaction is due only to the additional integrals of the quasiparticle
motion, other than the energy? Unfortunately, we see no possibility of giving a
sufficiently clear-cut and definite answer to such questions. Nonetheless, we wish to
point out the difficulties that would probably be encountered in attempts to reconcile
the obtained picture with the assumption concerning the Cooper phenomenon in nuclear
matter. It is precisely such a phenomenon which is characterized by a constant width
of the transition region of the statistical distribution, having the same value for all
the quasiparticles located near the Fermi boundary \footnote{A good illustration
is the well-known problem of an almost-ideal Fermi gas with weak attraction. Its
solution is frequently given in the momentum representation (see, for example, [6]).
It is difficult to visualize how the purely formal operation of the transition to
the $l$-representation could change the width of the transition band of the Fermi
distribution, and furthermore make it dependent on $l$ (see also footnote 6).}. However,
the presence of a constant component of the width of this band would lead to the
absence of the experimentally observed magic jumps of the binding energy of the
nucleons (see Appendix 2, and also footnotes 3 and 6). The available experimental
data therefore agree more readily with the simpler and natural hypothesis concerning
the character of the energy spectrum of unbounded nuclear matter as such. It seems
likely that it can represent a ``normal" Fermi liquid with an absolutely sharp Fermi
boundary for quasiparticles [7].

On the other hand, spherical nuclei of finite radius have a residual interaction of
the type characterized by (46). One must not forget, however, that it was actually
established only on the basis of data that pertained throughout to the nearest
vicinity of some magic nucleus. The experimental data give the impression that the
residual interaction of the considered type decreases gradually with increasing
distance from the magic nucleus, and becomes somehow restructured in final analysis.
The thermodynamics of the transition was considered in [8]. The fact that the phase
transition due to weakening and restructuring of the residual interaction causes
also the nuclear shape to become non-spherical can hardly be regarded as surprising.
It was already shown earlier (see [4]) that in the simplest scheme without interaction
the sphere is absolutely unstable in general for any number of particles.

We thank I.I. Gurevich, L.P. Kudrin, I.M. Pavlichenkov, G.A. Pik-Pichak, V.P. Smilga,
and K. A. Ter-Martirosyan for a discussion of the results.

\setcounter{equation}{0}

\renewcommand{\theequation}{A1.\arabic{equation}}

\section*{Appendix 1}

In the preceding paper [1] (see also [4]) we considered for concreteness a scheme
with a wall that is impermeable to the quasiparticles and is located at a distance
$R$ from the center of the nucleus. The roots of the wave equation of the free motion
of the particle in the spherical region, shown in Fig.~3 above, corresponded to such
a boundary condition. We shall show that this does not limit the generality of the
results pertaining to shell oscillations of the volume energy of the spherical nucleus.

The Bohr-Sommerfeld quantization rule [2] used to determine the eigenvalues is
\begin{equation}\label{1-1}
    \int_a^Rk_l(r)dr=\pi(n+\gamma).
\end{equation}
Here
\begin{equation}\label{1-2}
    k_l(r)=\sqrt{k^2-\tilde{l}^2/r^2},\qquad \tilde{l}=l+1/2,
\end{equation}
and the lower limit of integration is due to the centrifugal barrier; since the
nuclear matter is homogeneous, the wave number $k$ is constant in the interval region.
The additional phase $\gamma$ depends on the properties of the true structure of the
transition layer on the surface of the nucleus. We change over to the dimensionless
variable $kr = \rho'$:
\begin{equation}\label{1-3}
    \int_{\tilde{l}}^\rho \mathcal{R}(\rho')d\rho=\pi(n+\gamma),\quad
    \mathcal{R}=\sqrt{1-\tilde{l}^2/{\rho'}^2}.
\end{equation}
The Regge trajectories responsible for the oscillations are characterized by the relation
\begin{equation}\label{1-4}
    2n+l=p,\qquad p=2,3,4,\ldots,
\end{equation}
between the quantum numbers (see the Introduction and Fig.~3). Therefore
\begin{equation}\label{1-5}
    dn/d\tilde{l}=-1/2,\qquad d^2n/d\tilde{l}^2=0.
\end{equation}
We now differentiate the entire relation (\ref{1-3}) along the trajectory
\begin{equation}\label{1-6}
    \mathcal{R}(\rho)\frac{d\rho}{d\tilde{l}}-\tilde{l}\int_{\tilde{l}}^\rho
    \frac{d\rho'}{{\rho'}^2\mathcal{R}(\rho')}=\pi\left(\frac{dn}{d\tilde{l}}
    +\frac{d\gamma}{d\tilde{l}}\right).
\end{equation}
Elementary integration yields
\begin{equation}\label{1-7}
    \mathcal{R}(\rho)\frac{d\rho}{d\tilde{l}}-\arccos\frac{\tilde{l}}{\rho}=-\frac\pi2
    +\pi\frac{d\gamma}{d\tilde{l}}.
\end{equation}
It follows therefore that the derivative $d\rho/dl$ vanishes at $\tilde{l} = 0$ (the last
term in the right-hand side, which is due to the differentiation of the phase increment,
does not affect the result (see (\ref{1-9}) below). In the second differentiation of
the Bohr-Sommerfeld formula, we omit the terms that are known to vanish at the extremum:
\begin{equation}\label{1-8}
    \mathcal{R}(\rho)\frac{d^2\rho}{d\tilde{l}^2}+\frac1{\rho\mathcal{R}(\rho)}
    =\pi\frac{d^2\gamma}{d\tilde{l}^2}.
\end{equation}
We now take into account the fact that the phase $\gamma$ is determined by the actual
details of the nuclear interactions in the nearest vicinity of the upper limit of
integration. Any characteristic that influence the results can depend here on the
angular momentum $\tilde{l}$ only in the combination $\tilde{l}^2$. Therefore
\begin{equation}\label{1-9}
    \frac{d\gamma}{d\tilde{l}}=2\tilde{l}\frac{d\gamma}{d\tilde{l}^2},\quad
    \frac{d^2\gamma}{d\tilde{l}^2}=2\frac{d\gamma}{d\tilde{l}^2}\sim\frac1{\rho^2}.
\end{equation}
Consequently, at the maximum point $l = 0$ we have
\begin{equation}\label{1-10}
    d^2\rho/d\tilde{l}^2\cong -1/\rho,
\end{equation}
neglecting the terms $\sim l/\rho^2$. Thus, the form of Eq.~(1)
\begin{equation}\label{1-11}
    \Delta\rho\cong-(l+1/2)^2/2\rho
\end{equation}
does not depend on the details of the nuclear surface-layer structure. We determined
finally the ordinates of the extrema of the successive trajectories. At $\tilde{l}=l+1/2=0$
we have according to (\ref{1-4}) $n = p/2 + l/4$. Substituting in (\ref{1-4}), we get
\begin{equation}\label{1-12}
    \rho_{max}=\pi(\tfrac12 p+\gamma'),
\end{equation}
where $\gamma'=\gamma + l/4$. Replacing $\rho_{max}$ by $k_fR = \rho_f$, we see that
(\ref{1-12}) coincides in fact with the previously published (see [1], formula (19),
and footnote 9) ``quantization rule" for the magic values of this parameter.

\setcounter{equation}{0}

\renewcommand{\theequation}{A2.\arabic{equation}}

\section*{Appendix 2}

We consider here the special case $k = 0$, for which there is no asymptotic region
described by formulas of the type (23). The width of the band II of the smearing of
the Fermi distribution (see Fig. 4), which does not depend on $l$, will be denoted by
$g_0$. Formally, this requires that we put $k = 0$ in (17) and make the substitution
$g\rho_f\to g_0$. Substituting then in (34) and (35), we obtain after a simple integration
\begin{equation}\label{2-1}
    \widetilde{N}_1=\frac{\rho_f}{4\pi g_0}\sum_{\nu=1}^\infty
    \frac{\sin2\nu g_0\cos4\nu\rho_f}{\nu^2}.
\end{equation}

In the next application of formula (3), we transform also the trigonometric expressions
under the summation sign
\begin{equation}\label{2-2}
    E_1=-\frac{\eps\rho_f}{2\pi g_0}\left\{\EuFrak{L}\left(\rho_f+\frac{g_0}2\right)-
    \EuFrak{L}\left(\rho_f-\frac{g_0}2\right)\right\}.
\end{equation}
Here
\begin{equation}\label{2-3}
    \EuFrak{L}(\rho_f)=\frac14\sum_{\nu=1}^\infty\frac{\sin4\nu\rho_f}{\nu^2},
    \quad \frac{d\EuFrak{L}}{d\rho_f}=\EuFrak{M}(\rho_f).
\end{equation}
Since the expression for $\EuFrak{M}(\rho_f)$ (see formula (28) above, and also [1])
is characterized by a jump of the first derivative, only the second derivative of the
function $\EuFrak{L}(\rho_f)$ becomes discontinuous. Moreover, these singularities
(which are located in this case at the inflection points of the function $E_1(\rho_f)$)
are now shifted by a distance $\pm g_0/2$. The physical reason for this is seen
directly from Fig.~4. At the very point on the $\rho_f$ scale where the first
derivative would experience a discontinuity at $k = 2$, the expression (\ref{2-2})
has an ordinary analytic minimum. Thus, the nucleon binding energy $\eps$ experiences
no discontinuity anywhere.

\vspace{1cm}

\centerline{\bf Figures captions}

\bigskip

Fig.~1.

\bigskip

Fig.~2. Schematic diagram of the shell oscillations of the nuclear mass in the vicinity of the
magic number $N=126$.

\bigskip

Fig.~3. Roots of the Bessel functions with half-integer index on the $(l, \rho)$ diagram.

\bigskip

Fig.~4.

\bigskip

Fig.~5.

\bigskip

Fig.~6.

\end{document}